\documentclass[showpacs]{revtex4} %twocolumn
\usepackage{epsfig,amsmath,amssymb,graphicx,upgreek,textcomp}
\usepackage{natbib}
\usepackage[english]{babel} %

\begin{document}

\vspace{0mm}
\title{Coherence, broken symmetry and nondissipative motion of a
quantum oscillator} %
\author{Yu.M. Poluektov}
\email{yuripoluektov@kipt.kharkov.ua (y.poluekt52@gmail.com)} %
\affiliation{National Science Center ``Kharkov Institute of Physics and Technology'', 61108 Kharkov, Ukraine} %

\begin{abstract}
On the example of a quantum oscillator the connection of the
dynamical coherent state with the phase symmetry breaking and the
existence of the nondissipative motion is considered. In
multiparticle systems of interacting particles similar states
manifest themselves as superfluidity and superconductivity.
\newline%
{\bf Key words}: %
quantum oscillator, coherent states, broken phase symmetry,
anomalous and normal averages, pair correlations, superfluidity,
superconductivity
\end{abstract}
\pacs{03.65.\,--\,w, 05.30.Jp, 05.70.\,--\,a, 42.50.Ar, 63.70.\,--\,h, 64.60.De, 67.10.Fj} %

\maketitle

\vspace{0mm}
\section{Introduction}\vspace{-0mm} %\cite{}
For understanding the properties of multiparticle systems, exactly
solvable problems play an important role, in particular, the ideal
gas and harmonic quantum oscillator models. The dynamical coherent
state (DCS) of a quantum oscillator was considered as early as by
Schr\"{o}dinger at the dawn of quantum mechanics \cite{EShr,LL}. A
quantum oscillator can be considered as a kind of the simplest model
of a solid body \cite{HH}. The representation of coherent states
(CS) is widely used in the study of various quantum systems
\cite{MM,KS,AP,MV}.

In this paper, we consider a harmonic quantum oscillator in the
dynamical coherent state. It is paid attention that the peculiarity
of such a state is that it has a nondissipative internal motion with
a non-zero average momentum, and at the same time the symmetry with
respect to the phase transformation proves to be broken. It is noted
that the transition to the coherent state with broken phase symmetry
leads to the existence of nondissipative flows, which in Bose
systems manifest themselves as superfluidity, and in charged Fermi
systems as superconductivity.

\section{``Normal'' and coherent states of a quantum oscillator}\vspace{-0mm}
The Hamiltonian of a quantum oscillator \cite{LL}
\begin{equation} \label{01}
\begin{array}{l}
\displaystyle{%
  H=\frac{p^2}{2M}+\frac{M\omega^2x^2}{2}
}%
\end{array}
\end{equation}
can be represented in terms of non-selfadjoint creation $a^+$ and
annihilation $a$ operators, satisfying the commutation relation
$\big[\,a,a^+\big]\equiv aa^+-a^+a=1$. The coordinate and momentum
operators are determined through these operators by the relations
\begin{equation} \label{02}
\begin{array}{l}
\displaystyle{%
  x=\sqrt{\frac{\hbar}{2M\omega}}\,\big(a^++a\big), \qquad %
  p\equiv -i\hbar\frac{d}{dx}=i\sqrt{\frac{M\hbar\omega}{2}}\,\big(a^+-a\big), %
}%
\end{array}
\end{equation}
and the Hamiltonian (1) takes the known form
\begin{equation} \label{03}
\begin{array}{l}
\displaystyle{%
  H=\hbar\omega\bigg(a^+a+\frac{1}{2}\bigg). %
}%
\end{array}
\end{equation}
It is invariant with respect to the phase transformation
\begin{equation} \label{04}
\begin{array}{l}
\displaystyle{%
  a\rightarrow a'=ae^{i\alpha}, \quad %
  a^+\rightarrow a'^+=a^+e^{-i\alpha} %
}%
\end{array}
\end{equation}
or transformation of the coordinate and momentum operators
\begin{equation} \label{05}
\begin{array}{ll}
\displaystyle{%
  x\rightarrow x'=-\frac{p}{M\omega}\sin\alpha + x\cos\alpha, %
}\vspace{3mm}\\ %
\displaystyle{\hspace{0mm}%
  p\rightarrow p'=p\cos\alpha + M\omega x\sin\alpha, %
}%
\end{array}
\end{equation}
where $\alpha$ is an arbitrary real number. The eigenstates of the
Hamiltonian (3) $H|n\rangle=\varepsilon_n|n\rangle$ with energy
$\varepsilon_n\equiv\hbar\omega(n+1/2)$ are characterized by
integers $n=0,1,2,\ldots$. The time dependence of vectors in states
with fixed energy has the form
$|n,t\rangle=e^{-i\frac{\varepsilon_n}{\hbar}t}|n\rangle$. The
actions of the operators $a^+,\,a$ on the state vector are given by
the known relations $a^+|n\rangle=\sqrt{n+1}\,|n+1\rangle$,
$a|n\rangle=\sqrt{n}\,|n-1\rangle$, $a^+a|n\rangle=n|n\rangle$. %
The ground (vacuum) state $|0\rangle$ is defined as a solution of
the equation $a|0\rangle=0$, and the vectors of excited states of
the oscillator are found as a result of the action of powers of the
operator $a^+$ on the ground state %
$|n\rangle=\frac{(a^+)^n}{\sqrt{n!}}|0\rangle$. %
In the coordinate representation, the wave functions of the
oscillator are expressed through the Hermite polynomials $H_n(x)$ \cite{LL}: %
\begin{equation} \label{06}
\begin{array}{ll}
\displaystyle{%
  \varphi_n(x)=\bigg(\frac{M\omega}{\pi\hbar}\bigg)^{\!1/4} %
  \frac{1}{\sqrt{2^nn!}}\,e^{-\frac{M\omega}{2\hbar}\,x^2} %
  H_n\bigg(x\sqrt{\frac{M\omega}{\hbar}}\,\bigg). %
}%
\end{array}
\end{equation}
In the stationary states of the oscillator, the average values of
the coordinate and momentum are equal to zero
$\overline{x}\equiv\langle n|x|n\rangle=0$, $\overline{p}\equiv\langle n|p|n\rangle=0$, %
and the product of coordinate and momentum fluctuations %
$I\equiv\sqrt{\big(\overline{x^2}-\overline{x}^2\big)\big(\overline{p^2}-\overline{p}^2\big)}$ %
for the $n$-th level is $I_n=\hbar\big(n+\frac{1}{2}\big)$. The
minimum value of the product of coordinate and momentum fluctuations
is achieved in the ground state $I_0=\hbar/2$. Such states of the
oscillator with a certain energy will be called ``normal''.

An arbitrary time-dependent state vector can be decomposed over the
complete system of eigenvectors of the oscillator
\begin{equation} \label{07}
\begin{array}{ll}
\displaystyle{%
  \big|\Phi(t)\big\rangle =\sum_{n=0}^\infty
  C_n\,e^{-i\frac{\varepsilon_n}{\hbar}t}\,|n\rangle.
}%
\end{array}
\end{equation}
Let us consider such a specific state in which the expansion
coefficients in (7) have the form
\begin{equation} \label{08}
\begin{array}{ll}
\displaystyle{%
  C_n=\frac{\chi^n}{\sqrt{n!}}\,e^{-\frac{|\chi\,|^2}{2}}, \qquad %
  \sum_{n=0}^\infty \big|C_n\big|^2=1,
}%
\end{array}
\end{equation}
where $\chi$ is an arbitrary complex number being independent of
time. In the case of expansion coefficients (8), the probability of
finding the oscillator in the state $|n\rangle$ is determined by the
Poisson distribution
$\big|C_n\big|^2=e^{-|\chi|^2}\frac{|\chi|^{2n}}{n!}$. Then the
state (7) takes the form
\begin{equation} \label{09}
\begin{array}{ll}
\displaystyle{%
  \big|\Phi_\chi(t)\big\rangle = e^{-\frac{|\chi\,|^2}{2}} %
  \sum_{n=0}^\infty \frac{\chi^n}{\sqrt{n!}}\,e^{-i\frac{\varepsilon_n}{\hbar}t}|n\rangle.%
}%
\end{array}
\end{equation}
At $\chi=0$ (9) coincides with the wave function of the oscillator
in the ground state. The vector (9) is an eigenstate of the
annihilation operator
$a\big|\Phi_\chi(t)\big\rangle=\chi(t)\big|\Phi_\chi(t)\big\rangle$
with the time-dependent eigenvalue $\chi(t)\equiv\chi e^{-i\omega t}$. %
Note that usually one considers the stationary coherent states \cite{MM,KS,AP,MV}: %
\begin{equation} \label{10}
\begin{array}{ll}
\displaystyle{%
  |\chi\rangle\equiv\big|\Phi_\chi(0)\big\rangle = e^{-\frac{|\chi\,|^2}{2}} %
  \sum_{n=0}^\infty \frac{\chi^n}{\sqrt{n!}}\,|n\rangle.%
}%
\end{array}
\end{equation}
The dynamical coherent state can be obtained by the action of the
operator $U(t)=\exp\!\big[\!-i\omega t\big(a^+a+\frac{1}{2}\big)\big]$ %
on the stationary CS. It is important to emphasize that the
stationary CS is not a solution of the stationary Schr\"{o}dinger
equation, whereas the DCS (9) is an exact solution of the
nonstationary Schr\"{o}dinger equation. Consideration of the
nonstationary coherent states is fundamentally important in the
study of systems in which nondissipative flows may exist. %

In the coordinate representation the function (9) takes the form
\begin{equation} \label{11}
\begin{array}{ll}
\displaystyle{%
  \Phi_\chi(x,t) = \bigg(\frac{M\omega}{\pi\hbar}\bigg)^{\!1/4}e^{-\frac{|\chi\,|^2}{2}}\, %
  e^{-i\frac{\omega t}{2}}\,e^{-\frac{M\omega}{2\hbar}\,x^2} %
  \sum_{n=0}^\infty\frac{\chi(t)^n}{\sqrt{2^n}n!}\,H_n\bigg(x\sqrt{\frac{M\omega}{\hbar}}\,\bigg). %
}%
\end{array}
\end{equation}
Using the formula for a sum of Hermite polynomials
\begin{equation} \label{12}
\begin{array}{ll}
\displaystyle{%
  \sum_{k=0}^\infty\frac{t^k}{k!}\,H_k(x)=e^{2xt-t^2}, %
}%
\end{array}
\end{equation}
we obtain the representation of function (11) in the form
\begin{equation} \label{13}
\begin{array}{ll}
\displaystyle{%
  \Phi_\chi(x,t) = \bigg(\frac{M\omega}{\pi\hbar}\bigg)^{\!1/4}e^{-\frac{|\chi\,|^2}{2}}\, %
  e^{-i\frac{\omega t}{2}+\frac{\chi^2(t)}{2}}\,e^{-\frac{M\omega}{2\hbar}\left(x-\chi(t)\sqrt{\frac{2\hbar}{M\omega}}\,\right)^{\!2} }. %
}%
\end{array}
\end{equation}
This function is a wave packet that does not diffuse with time.

\section{Difference between symmetries of ``normal'' and coherent states}\vspace{-0mm}
As noted, the Hamiltonian (3) is symmetric with respect to the phase
transformations (4). If the oscillator is in states with fixed
energy $\varepsilon_n\equiv\hbar\omega\big(n+\frac{1}{2}\big)$, then
the averages of only the phase-invariant operators can be non-zero
$\langle n|a^+a|n\rangle=n$, while the averages of the
phase-noninvariant operators $\langle n|a|n\rangle=\langle
n|a^+|n\rangle=0$ and $\langle n|a^2|n\rangle=\langle
n|a^{+2}|n\rangle=0$ are equal to zero. Thus, the symmetry of the
averages in the normal states coincides with the symmetry of the
Hamiltonian.

In the coherent states, both the normal averages from the
phase-invariant creation and annihilation operators and the
anomalous averages from the phase-noninvariant operators turn out to
be non-zero:
\begin{equation} \label{14}
\begin{array}{ccc}
\displaystyle{%
  \big\langle\Phi_\chi(t)\big|a^+a\big|\Phi_\chi(t)\big\rangle=|\chi|^2, %
}\vspace{3mm}\\ %
\displaystyle{\hspace{0mm}%
  \big\langle\Phi_\chi(t)\big|a^+\big|\Phi_\chi(t)\big\rangle=\chi^*(t), \quad %
  \big\langle\Phi_\chi(t)\big|a\big|\Phi_\chi(t)\big\rangle=\chi(t), %
}\vspace{3mm}\\ %
\displaystyle{\hspace{0mm}%
  \big\langle\Phi_\chi(t)\big|a^{+2}\big|\Phi_\chi(t)\big\rangle=\chi^{*2}(t), \quad %
  \big\langle\Phi_\chi(t)\big|a^2\big|\Phi_\chi(t)\big\rangle=\chi^2(t). %
}%
\end{array}
\end{equation}
In this case, the symmetry of the anomalous averages turns out to be
lower than the symmetry of the Hamiltonian (3). Such states are
called states with spontaneously broken phase symmetry.

The average values of the coordinate and momentum in the coherent
state, in contrast to the normal state, are non-zero and depend on time: %
\begin{equation} \label{15}
\begin{array}{ll}
\displaystyle{%
  \overline{x}(t)=\big\langle\Phi_\chi(t)\big|\,x\big|\Phi_\chi(t)\big\rangle= %
  \sqrt{\frac{\hbar}{2M\omega}}\,\big(\chi^*(t)+\chi(t)\big), %
}\vspace{3mm}\\ %
\displaystyle{\hspace{0mm}%
  \overline{p}(t)=\big\langle\Phi_\chi(t)\big|\,p\,\big|\Phi_\chi(t)\big\rangle= %
  i\sqrt{\frac{M\hbar\omega}{2}}\,\big(\chi^*(t)-\chi(t)\big). %
}%
\end{array}
\end{equation}
It follows that $\dot{\overline{x}}(t)=\overline{p}(t)/M$,
$\dot{\overline{p}}(t)=-M\omega^2\,\overline{x}(t)$, and the average
values of both coordinate and momentum satisfy the classical
oscillator equation
$\ddot{\overline{x}}(t)+\omega^2\overline{x}(t)=0$,
$\ddot{\overline{p}}(t)+\omega^2\overline{p}(t)=0$. In contrast to
the stationary state in the normal state of the oscillator, where
the average momentum is zero, in the dynamical coherent state the
average momentum is non-zero and oscillates in time. Therefore, in
the dynamical coherent state there exists a nondissipative internal
motion, which is analogous to states of multiparticle systems with
nondissipative flows. The wave function of the coherent state (13)
can be expressed through the averages $\overline{x}(t), \overline{p}(t)$: %
\begin{equation} \label{16}
\begin{array}{ll}
\displaystyle{%
  \Phi_\chi(x,t) = \bigg(\frac{M\omega}{\pi\hbar}\bigg)^{\!1/4}\,
  e^{-i\,\left(\frac{\omega t}{2}+\frac{\overline{p}(t)\overline{x}(t)}{2\hbar}\right)}\, %
  e^{i\frac{\overline{p}(t)}{\hbar}x}\, %
  e^{-\frac{M\omega}{2\hbar}(x-\overline{x}(t))^2}. %
}%
\end{array}
\end{equation}
In this form, this wave function was first obtained by
Schr\"{o}dinger \cite{EShr,LL}.

The averages over the coherent state from the squares of coordinate
and momentum have the form
\begin{equation} \label{17}
\begin{array}{ll}
\displaystyle{%
  \overline{x^2}(t)=\big\langle\Phi_\chi(t)\big|\,x^2\big|\Phi_\chi(t)\big\rangle= %
  \frac{\hbar}{2M\omega}\big(\chi(t)^{*2}+\chi(t)^2+2|\chi|^2+1\big), %
}\vspace{3mm}\\ %
\displaystyle{\hspace{0mm}%
  \overline{p^2}(t)=\big\langle\Phi_\chi(t)\big|\,p^2\big|\Phi_\chi(t)\big\rangle= %
  -\frac{M\hbar\omega}{2}\big(\chi(t)^{*2}+\chi(t)^2-2|\chi|^2-1\big). %
}%
\end{array}
\end{equation}
Taking into account (15) and (17), we find for the uncertainty relation %
\begin{equation} \label{18}
\begin{array}{ll}
\displaystyle{%
  I\equiv\sqrt{\big(\overline{x^2}-\overline{x}^2\big)\big(\overline{p^2}-\overline{p}^2\big)}=\frac{\hbar}{2}. %
}%
\end{array}
\end{equation}
As was shown already by Schr\"{o}dinger \cite{EShr,LL}, the
uncertainty $I$ (18) in the coherent state is minimal. For the
normal state of the oscillator, the uncertainty is minimal only in
the ground state.

In the coherent state, the energy of the oscillator is not precisely
determined. The quantum-mechanical average of energy in the
dynamical coherent state is independent of time and can be expressed
through the averages of the squares of coordinate and momentum (17)
\begin{equation} \label{19}
\begin{array}{cc}
\displaystyle{%
  \varepsilon(\chi)=\hbar\omega
  \Big(\big\langle\Phi_\chi(t)\big|a^+a\big|\Phi_\chi(t)\big\rangle+\frac{1}{2}\,\Big) = %
  \hbar\omega\Big(|\chi|^2+\frac{1}{2}\,\Big)= %
}\vspace{3mm}\\ %
\displaystyle{\hspace{0mm}%
  =\frac{\overline{p^2}(t)}{2M}+\frac{M\omega^2}{2}\,\overline{x^2}(t). %
}%
\end{array}
\end{equation}
Thus, the dynamical coherent state is the state with a constant
average energy, in which there exists an undamped motion and the
phase symmetry is broken.

\section{Discussion. Conclusions}\vspace{-0mm}
The main feature of superfluid and superconducting systems is the
possibility of stationary flows of mass or charge in them, which can
exist arbitrarily long time without attenuation. The transition from
the normal state to the superfluid or superconducting state is a
phase transition. According to the general theory of phase
transitions \cite{LL2}, a phase transition to an ordered state
should be accompanied by the appearance of a new characteristic --
the order parameter. For a long time it was unclear what is the
order parameter in superconducting or superfluid transitions. Even
before the discovery of superfluidity, L.V. Shubnikov \cite{ShK}
suggested a hypothesis that the phase transition He I -- He II is
accompanied by an ordering similar to the transition from a liquid
to a crystalline state. However, after superfluidity was discovered
\cite{KL,AM}, it became unclear how such ordering is able to support
nondissipative flows. The correct form of the order parameter in the
framework of the phenomenological description was proposed by
Ginzburg and Landau \cite{GL}. They used the complex macroscopic
wave function as an order parameter for superconductors. Such an
order parameter allows to construct an expression for the
macroscopic density of the nondissipative flow, similar to the way
the probability flux density is constructed in quantum mechanics.
Since the macroscopic wave function, like the wave function in
quantum mechanics, is defined up to an arbitrary phase factor, then
such a state is called a state with broken phase symmetry. In a
simple version for superconductors, a microscopic justification of
the Ginzburg-Landau theory on the basis of the BCS theory \cite{BCS}
was given by Gorkov \cite{GL2}. He showed that the existence of
superconducting properties is associated with the appearance of the
anomalous averages. The phenomenological Ginzburg-Landau approach
was extended by Ginzburg and Pitaevskii to a superfluid Bose liquid
\cite{GP}. The development of this theory was continued in works
\cite{GS,P1}. It should be noted that in the well-known Landau
theory of superfluidity \cite{L1,L2} the violation of phase symmetry
is also implicitly taken into account through the introduction of
the superfluid density associated with the modulus of the complex
order parameter, and the superfluid velocity determined by the phase
gradient. In Bose systems, a simple modeling of superfluidity at
zero temperature leads to the Gross-Pitaevskii equation \cite{Gr,Pit}. %
The macroscopic wave function in this theory is the coherent state \cite{P2}. %

For the macroscopic complex wave function to exist, the system must
have non-zero anomalous averages of the form $\langle
a_{k_1\sigma_1}a_{k_2\sigma_2}\rangle$ or $\langle
a_{k\sigma}\rangle$ that violate the phase symmetry. In this case,
the macroscopic wave function of the superconductor in the
$s$\,-\,state will be of the form %
$\Psi({\bf r}_1,{\bf r}_2)\sim\sum_{{\bf k}_1,{\bf k}_2}e^{-i({\bf
k}_1{\bf r}_1+{\bf k}_2{\bf r}_2)}\langle a_{{\bf
k}_1\uparrow}a_{{\bf k}_2\downarrow}\rangle$, %
and the macroscopic wave function of the Bose system of particles
with zero spin -- of the form %
$\Psi({\bf r})\sim\sum_{\bf k}e^{-i{\bf k}{\bf r}}\langle a_{{\bf k}}\rangle$.%

At present, as already seen from the cited works, it is quite clear
that in any physical system where the phenomena of superfluidity or
superconductivity exist, the phase symmetry must be necessarily
violated. At the same time, of course, each system also has
properties that are specific to it. Nevertheless, there appeared and
continues to appear a large number of works devoted to this problem,
where the violation of the phase symmetry is absent. In this case,
there cannot be states with equilibrium nondissipative flows \cite{P3}. %

However, the problem of establishing at the microscopic level the
connection between the coherence of the state, the violation of the
phase symmetry and the existence of the nondissipative flows in
multiparticle Fermi and Bose systems continues to remain actual. In
this paper, using a simple example of a quantum harmonic oscillator,
the connection is traced between the violation of the phase symmetry
and the existence of the macroscopic motion in the dynamical
coherent state. It is shown that, in contrast to the states of the
oscillator with a fixed energy, in which the average momentum is
equal to zero, in the dynamical coherent state it is different from
zero and oscillates according to the equation for a classical
oscillator. At the same time, in the DCS the phase-noninvariant
anomalous averages also turn out to be different from zero.

The author thanks A.A.\,Soroka for help in preparing the article.

\end{document}